\documentstyle [12pt, A4]{article}
\thicklines                     
\begin{document}
\rightline {CbNU-Th-950724}
\rightline {July, 1995}

\vspace{.0in}
\begin{center}
{\large\bf Non-Hermitian quantum canonical variables and the generalized ladder
operators}\\[.2in]
{W. S. l'Yi\footnote{wslyi@cbucc.chungbuk.ac.kr}}\\[.15in]
{\it Department of Physics\\
Ch'ungbuk National University\\
Ch'ongju, 360-763, Ch'ungbuk, Korea} \\[.5in]
\end{center}

\begin{center}
{\bf ABSTRACT}\\
\end{center}
\begin{quotation}
Quantum canonical transformations of the second kind and the non-Hermitian
realizations
of the basic canonical commutation relations are investigated with a special
interest in
the generalization of the conventional ladder operators.  The opeator ordering
problem is
shown to be resolved when the non-Hermitian realizations for the canonical
variables
which can not be measured simultaneously with the energy are chosen for the
canonical
quantizations.  Another merit of the non-Hermitian representations is that it
naturally
allows us to introduce the generalized ladder operators with which one can
solve eigenvalue
problems quite neatly.
\end{quotation}
\smallskip
PACS number(s): 03.65.Ca, 04.20.Fy, 04.60.Ds
\newpage

{\large\bf I. Introduction}\vspace{.10in}

Canonical transformations which include, as parts, both point transformations
and time
evolutions are not only theoretically but also practically important concepts
for
solving classical problems.  Constructions of practically useful quantum
versions of the
canonical transformations are, on the other hand, as elusive as solving the
quantum mechanical
equations of motions themselves.  Inspired by the beauty of the classical
canonical
transformations there appeared several interesing attempts.
The widely known endeavor relied on the Hamiltonian path integral qunatization
techniques \cite{hamiltonian_path_al}.  The virtue of this formalism is
that all the physical quantities are pure numbers and there are no operator
ordering problems.
But there are still pitfalls in this approach.  One is that
the canonical momenta $p_i$ and $p_{i+1}$ of a path integral at time slices $t$
and
$t+\epsilon$ are unrelated.  In other word $q_i(p_{i+1} -p_i)$ is longer
$\epsilon q_i\dot{p_i},$ and all the complications arise.

There are another attempts which may possibly circumvent this problem.
The ``effective generating technique'' by some authors \cite{effective} is one
of
the proposals.  In this approach the
quantum generating function of a canonical transformation is written in terms
of a
series expansion in powers of $\hbar$ whose leading order term is the
corresponding
classical one.  On the other hand Anderson extended unitary canonical
transformations to non-unitary ones \cite{non_Hermitian_QCT}.
Even though it is quite general it lacked clear classical analogy.
To improve this weak point another proposal \cite{lyi} based on the more
traditional ``mixed
matrix element technique'' \cite{mixed} is presented.  In that paper it is
shown that to get
useful quantum canonical transformations one should allow non-Hermitian
representations for
the various canonical variables.  It is clear that for the dynamical variables
which can not
be measured with energy simultaneously one may freely choose non-Hermitian
representations.
In this non-Hermitian operator technique the classical analogy is preserved
upon quatization,
and higher $\hbar$ power terms of the effective generating function are
interpreted as
non-Hermitian mordifications to the Hamiltonian operator corresponding to the
classically
transformed Hamiltonian.
But this operator technique of canonical transformation has both advantages and
disadvantages.
The facts that this surmounts the operator ordering problems by the concept of
``well-orderedness'' and that whenever there is a classically useful generating
function
there is a high probability to solve the Schr\"odinger equation, cause us to
show favor to
this approach.
In that paper various kinds of the quantum canonical
transformations inspired by the classical counter parts are introduced.
One of the disadvantages, on the other hand, is that it is usually difficult to
solve the old
quantum canonical variables $(q_r,p_s)$ in terms of the new ones $(Q_i,P_j),$
thus
prohibiting us to write the new Hamiltonian $K=H+{\partial F^\dagger\over
\partial t}$ in
terms of $(Q_i,P_j).$  But there are still quite large portions of quantum
canonical transformations which are practically useful.

In this paper we consider quantum canonical transformations of the second kind
with a special
emphasis on the point transformations in relation to the generalized ladder
operators.
We show that non-Hermitian representations of the canonical commutation
relations
\begin{equation}
[q_r,p_s]=i\delta_{rs},\;\;\;
[q_r,q_s]=0,\;\;\;[p_r,p_s]=0,\label{canonical_commutations}
\end{equation}
greatly simplify the quantization process of classical systems.  It is also
shown that
the generalized ladder operators which are naturally associated with the
non-Hermitian
canonical variables allow us to solve eigenvalue problems quite elegantly.

General ideas on the quantum canonical transformations of the second kind with
relation to
coordinate reparametrizations are presented in the next section.  The
construction of
the generalized ladder operators and a systematic way of solving eigenvalue
equations are
discussed in Sec.~III.   Some applications are shown in Sec.~IV.  The
conclusion is given
in Sec.~V.
\vspace{.10in}

{\large\bf II. Quantum canonical transformations}

In this paper we follow our previous notations which dealt on the general idea
of
the canonical transformations \cite{lyi}.
Let $|q'\rangle=|q'_1,\dots,q'_f\rangle$ be a simultaneous eigenket of
observables $q_r,$ $r=1, \dots, f,$ such that
\begin{eqnarray}
 q_r|q'\rangle &=& q'_r|q'\rangle, \\
 \langle q' | q'' \rangle  &=& {1\over\rho(q')} \delta(q'-q''),\\
 {\bf 1} &=& \int d^fq'\, |q'\rangle \rho(q') \langle q'|,
\end{eqnarray}
where we use the convention that various eigenvalues of an observable $q_r$ are
denoted by
attaching primes such as $q'_r,$ $q''_r,$ etc.
To investigate the general properties of coordinates transformations we
introduce a set of
functions $f_i(q_1',\dots, q_f'),$ $i=1,\dots,f,$ such that
\begin{equation}
\det\left( {\partial f_i \over \partial q'_r} \right) \neq 0.
\end{equation}
Using $|q'\rangle$ and a generating function given by
\begin{equation}
F(q_1',\dots,q_f', P_1',\dots,P_f') = \sum_{i=1}^f f_i(q') P'_i,
\label{gen_fun}
\end{equation}
we define another set of kets $|P'\rangle$ by
\begin{equation}
\langle q'|P'\rangle=e^{\textstyle{iF(q',P')}}.
\label{canonical_transformation}
\end{equation}
For the reason of simplicity we consider only real functions $f_i.$
{}From the completeness of $|q'\rangle$ it is easy to prove, after straight
forward
computations, that $|P'\rangle$ also forms a complete set and that the
$q$-space and the
$P$-space scale density functions $\rho(q')$ and $\rho(P')$ are
\begin{equation}
\rho(q')=\left|\det\left( {\partial f_i \over \partial
q'_r}\right)\right|,\;\;\;\label{rho}
\rho(P')={1\over 2\pi}.
\end{equation}
The completeness of $|P'\rangle$ allows us to define Hermitian operators
$P_i,$ $i=1, \dots, f,$ such that
\begin{equation}
P^{}_i|P'\rangle = P'_i|P'\rangle.
\end{equation}

The physical meaning of $P_i$ will become transparent when we interpret $q'_r
\rightarrow P'_i$
as a part of a canonical transformation of the second kind corresponding to the
classical
generating function (\ref{gen_fun}).
The {\it well-ordered} generating operator which satisfies
$\langle q'|F(q,P)|P'\rangle = \langle q'|F(q',P')|P'\rangle$ is
\begin{equation}
F(q, P) = \sum_{i=1}^f f_i(q) P_i.
\end{equation}
Canonical operators $p_r,$ and $Q_i,$ which are defined by
\begin{eqnarray}
\langle q'|p_r|P'\rangle &=& -i{\partial\over\partial q'_r} \langle
q'|P'\rangle,\\
\langle P'|Q_i|q'\rangle &=& i{\partial\over\partial P'_i} \langle
P'|q'\rangle,
\end{eqnarray}
have following forms
\begin{eqnarray}
  p_r &=& {\partial F\over\partial q_r}
       = \sum_{i=1}^f {\partial f_i\over\partial q_r} P_i,\label{p}\\
  Q_i &=& {\partial F^{\dagger}\over\partial P_i} = f_i(q).\label{Q}
\end{eqnarray}
The equation (\ref{Q}) shows that the canonical transformation
(\ref{canonical_transformation}) corresponds to a reparametrization of the
coordinates.
It is important to notice that not only $P_i,$ which from the very definition,
but also
$Q_i,$ as it can be seen from (\ref{Q}), are Hermitian operators.
But the Hermitian conjugation of $p_r$ is
\begin{equation} \label{p_dag}
p_r^{\dagger}={1\over \rho(q)}p_r^{}\rho(q).
\end{equation}
It is true that instead of the non-Hermitian $p_r$ one may choose a Hermitian
combination
${1\over 2}(p_r^{} +p_r^\dagger).$  But it is obvious that it is not imperative
to use
Hermitian representations even for the canonical variables which can not be
measured
with energy simultaneously.  It will be soon clear that the freedom of choosing
the
non-Hermitian representations for some canonical variables makes us more
versatile in various
ways.

As an application of this idea consider a {\it Hermitian} Hamiltonian
$H = {1\over 2} \sum_i P^2_i  + V(Q).$
For later conveniences we rewrite this as
\begin{equation}
H = {1\over 2} P^{\dagger}_i P_i^{}  + V(Q), \label{H}
\end{equation}
where we used the summation convention for the repeated indices.
Inverting (\ref{p}) we have
\begin{equation}
P_i={\partial q_r \over \partial Q_i} p_r.
\end{equation}
The corresponding $q$-space Hamiltonian is
\begin{equation}
H = {1\over 2} p_r^{\dagger}{\partial q_r \over \partial Q_i}
  {\partial q_s \over \partial Q_i} p_s^{} + V. \label{hamiltonian_ugly}
\end{equation}
To simplify this equation we define a metric tensor $g_{ij}(Q)=\delta_{ij}$ in
such a way
that $ds^2=g_{ij}dQ^i dQ^j$ is the invariant line element of coordinate
transformations.
The $q$-space metric tensor is, then,
\begin{equation}
g^{rs}(q) = {\partial q_r \over \partial Q_i}{\partial q_s \over \partial
Q_i}.\label{g}
\end{equation}
It follows that the Hamiltonian operator (\ref{hamiltonian_ugly}) corresponding
to
the classical Hamiltonian $H = {1\over 2} p_r g^{rs} p_s + V$ simplifies to
\begin{equation}
H = {1\over 2} p_r^{\dagger} g^{rs} p_s^{} + V. \label{H_operator}
\end{equation}
Hamiltonian operators therefore can be unambiguously---that is, with no
ordering
ambiguity---constructed from the classical Hamiltonians as long as the
non-Hermitian forms
of canonical variables are used.

At this point we would like to emphasize that the $q$-space scale density
function (\ref{rho})
obtainded from the completenesses of $|q'\rangle$ and $|P'\rangle$ is the same
as that of
the one obtained from (\ref{g})!  It is reflected in the fact that the
canonical
transformation (\ref{canonical_transformation}) is unitary and that
(\ref{H_operator})
is Hermitian \cite{lyi}.

As a nontrivial illustration of this idea consider a free symmetrical top
described by
the following Hamiltonian function
\begin{equation}
H = {p_\theta^2 \over 2I_1} + { (p_\phi - p_\chi\cos\theta)^2 \over 2I_1
\sin^2\theta}
	+{p^2_\chi \over 2 I_3}, \label{H_top}
\end{equation}
where $\theta,$ $\phi,$ and $\chi$ are the Euler angles describing the
orientation of
the symmetrical top, and $I_1$ and $I_3$ denote the moment of inertia along the
principal axes.  The ranges of the Euler anlges are
\begin{equation}
0\leq \theta < \pi,\;\;\; 0\leq \phi < 2\pi,\;\;\; 0\leq \chi < 2\pi.
\end{equation}
The metric tensor read off from (\ref{H_top}) is
\begin{equation}
g^{rs} = \left(
  \begin{array}{ccc}
  {1\over I_1} & 0 & 0  \\
   0 & {1\over  I_1 \sin^2\theta} & -{\cos\theta \over I_1 \sin^2\theta} \\
   0 & -{\cos\theta \over I_1 \sin^2\theta} &
   			{1\over I_3} + {\cos^2\theta\over I_1\sin^2\theta} \end{array} \right),
\end{equation}
thus allowing us to write the scale density function $\rho(\theta\phi\chi)$ of
the
Euler angles,
\begin{equation}
\rho=\sqrt{ \det g_{rs} } = \sqrt{I_1^2I_3}\,\sin\theta.
\end{equation}
It is clear from (\ref{p_dag}) that $p_\phi$ and $p_\chi$ which are defined by
\begin{equation}
p_\phi =  -i{\partial \over \partial \phi}, \;\;\;
p_\chi =  -i{\partial \over \partial \chi}, \label{pp_op}
\end{equation}
are Hermitian, but the Hermitian conjugation of $p_\theta$ is
\begin{equation}
p^\dagger_\theta={1\over \sin\theta} p_\theta^{} \sin\theta.
\end{equation}
The Hamiltonian operator is therefore
\begin{equation}
H={p^\dagger_\theta p_\theta^{} \over 2I_1}
  + { (p_\phi - p_\chi\cos\theta)^2 \over 2I_1 \sin^2\theta}
  +{p^2_\chi \over 2 I_3}, \label{H_top_op}
\end{equation}
or, more explicitly,
\begin{equation}
H = -{1\over 2I_1}{1\over\sin\theta}{\partial\over\partial \theta}
		\sin\theta{\partial\over\partial \theta}
	-{1\over 2I_1\sin^2\theta}
	 ( {\partial\over\partial\phi} - {\partial\over\partial\chi}\cos\theta)^2
	-{1\over 2I_3}{\partial^2\over \partial \chi^2}
\end{equation}
which of course coincides with the known result \cite{fowler}.
\vspace{.10in}

{\large\bf III. Generalized ladder operators}

The non-Hermitian realization of canonical variables and the general form
(\ref{H_operator})
of the quantum mechanical Hamiltonian inspires us to introduce the following
concept of
the generalized ladder operators which is a generalization of the operator
factorization
concept of differential equations discussed by Infeld and Hull \cite{infeld}.
Suppose ${\cal E}$ is a Hibert space which can be decomposable into subspaces
${\cal E}(l),$
\begin{equation}
{\cal E} = \bigoplus_{l}{\cal E}(l),
\end{equation}
in the way that in the subspace ${\cal E}(l),$ which is usually an invariant
subspace of a
symmetry group ${\cal G}$ of the Hamiltonian, $H$ is effectively $H(l).$
In addition to this suppose that $H(l)$ can be written as
\begin{eqnarray}
H(l) &=& a(l)^\dagger a(l) + \nu(l)\\
     &=& a(l+1)a(l+1)^\dagger + \nu(l+1) +\epsilon(l+1),
\end{eqnarray}
where $\nu$ and $\epsilon$ are real numbers.  Then one obtains following two
equations
\begin{eqnarray}
H(l-1)a(l) &=& a(l)\{ H(l) + \epsilon(l) \}, \label{H(l-1)} \\
H(l+1)a(l+1)^\dagger &=& a(l+1)^\dagger\{ H(l) - \epsilon(l+1)
\label{H(l+1)}\}.
\end{eqnarray}
Whenever these hold solutions of the eingenvalue problems are automatic.

Suppose that $|El\tau\rangle$ which belonging to ${\cal E}(l)$ is an eigenstate
of
$H$ with the eigenvalue $E.$  $\tau$ is a set of any other quantum numbers
which are
irrelevant in this consideration.
Multiplying this eigenket to the right of (\ref{H(l-1)}) one get
\begin{equation}
H(l-1)a(l)|El\tau\rangle = \{E+\epsilon(l)\}a(l)|El\tau\rangle,
\end{equation}
showing that $a(l)|El\tau\rangle$ which belonging to ${\cal E}(l-1)$ is an
eigenstate
of $H$ with the eigenvalue $E+\epsilon(l).$  This means that $a(l)$ is a
descending ladder operator which raises energy by $\epsilon(l),$
\begin{equation}
\cdots \stackrel{a(l-1)}{\longleftarrow} {\cal E}(l-1)
	   \stackrel{a(l)}{\longleftarrow} {\cal E}(l)
	   \stackrel{a(l+1)}{\longleftarrow} {\cal E}(l+1)
	   \stackrel{a(l+2)}{\longleftarrow} \cdots
\end{equation}
On the other hand, when one uses (\ref{H(l+1)}), one gets followng relation
\begin{equation}
H(l+1)a(l+1)^\dagger|El\tau\rangle =
\{E-\epsilon(l+1)\}a(l+1)^\dagger|El\tau\rangle,
\end{equation}
showing that $a(l+1)^\dagger|El\tau\rangle$ which belonging to ${\cal E}(l+1)$
is an
eigenstate of $H$ with the eigenvalue $E-\epsilon(l+1).$  This means that
$a(l+1)^\dagger$
is an ascending ladder operator which lowers energy by $\epsilon(l+1),$
\begin{equation}
\cdots \stackrel{a(l-1)^\dagger}{\longrightarrow} {\cal E}(l-1)
	   \stackrel{a(l)^\dagger}{\longrightarrow} {\cal E}(l)
	   \stackrel{a(l+1)^\dagger}{\longrightarrow} {\cal E}(l+1)
	   \stackrel{a(l+2)^\dagger}{\longrightarrow} \cdots
\end{equation}

With the help of these operators we have following two schems for solving
eigenvalue
equations.

\noindent{\bf Case-I. Strings of ascending states.}\\
This is the case when there is a lower limit on the descension.
In this case let $n=l_{min}$ and solve
\begin{equation}
a(n)|{\rm lowest\;\; state}\rangle = 0.
\end{equation}
Non-degeneracy of the solution gaurentees the non-degeneracy of the higher
states.
Normalizing this ket we recast this as $|nl_{min}\tau\rangle$ and define
\begin{eqnarray}
E_{n,l_{min}} &=& \nu(n), \\
E_{n,l} &=& E_{n,l_{min}} - \sum_{k=l_{min}+1}^l \epsilon(k),\;\;\; l >
l_{min}.
\end{eqnarray}
As long as $E_{n,l-1} \ne \nu(l) + \epsilon(l),$ the vector
\begin{equation}
|nl\tau\rangle = {a(l)^\dagger \over \sqrt{E_{n,l-1} - \nu(l) - \epsilon(l)}}
                 |n,l-1,\tau\rangle \label{ket1}
\end{equation}
is a normalized eigenstate of $H$ with the eigenvalue $E_{nl}.$

\noindent{\bf Case-II. Strings of descending states.} \\
This is the case when there is an upper limit on the ascension.
In this case let $n=l_{max}$ and solve
\begin{equation}
a(n+1)^\dagger|{\rm highest\;\; state}\rangle = 0. \label{caseii-a}
\end{equation}
Properly normalizing this ket we rewrite this as $|nl_{max}\tau\rangle$ and
define
\begin{eqnarray}
E_{n,l_{max}} &=& \nu(n+1) + \epsilon(n+1), \\
E_{n,l} &=& E_{n,l_{max}} + \sum_{k=l+1}^{l_{max}} \epsilon(k),\;\;\; l <
l_{max}.
\end{eqnarray}
Then, as long as $E_{n,l+1} \ne \nu(l+1),$ the vector
\begin{equation}
|nl\tau\rangle = {a(l+1) \over \sqrt{E_{n,l+1} - \nu(l+1)}}
                 |n,l+1,\tau\rangle \label{ket2}
\end{equation}
is a normalized eingenstate of $H$ with the eigenvalue $E_{n,l}.$

There is a third class of cases which have limitations both on the ascensions
and descensions,
but it is not necessary to consider them seperately.  It can be considered as a
special case
of I or II.
\vspace{.10in}

{\large\bf IV. Applications of the generalized ladder operators}

Consider a three dimensional
spherically symmetric system described by the following Hamiltonian
\begin{equation}
H = {p_r^\dagger p_r^{} \over 2} + { {\bf L}^2 \over 2r^2 } + V(r),
\end{equation}
where $p^{\dagger}_r = {1\over r^2}p_r r^2$ and ${\bf L}^2$ is the angular
momentum operator
\begin{equation}
{\bf L}^2 = p_\theta^\dagger p_\theta^{}
 + {p_\phi^2 \over \sin^2\theta}. \label{L^2}
\end{equation}
Here the Hermitian conjugation of $p_\theta$ is ${1\over \sin\theta}p_\theta
\sin\theta.$
Since this operator is invariant under the action of the $SO(3)$
transformations the
Hilbert space ${\cal E}$ can be decomposed in terms of the eigenvector space
${\cal E}(l)$ of the angular momentum operator,
\begin{equation}
{\cal E} = \bigoplus_{l=0}^\infty {\cal E}(l).
\end{equation}
(Using our generalized ladder operator technique one may solve the eigenvalue
problem for
${\bf L}^2$ given by (\ref{L^2}).  This interesting digression is presented in
the Appendix.)
The Hamiltonian $H$ in ${\cal E}(l)$ is
\begin{equation}
H(l) = {p_r^\dagger p_r^{} \over 2} + { l(l+1) \over 2r^2 } + V(r).
\end{equation}

To be more specific, consider a three dimensional Harmonic oscillator described
by the
potential $V(r)={1 \over 2}\omega^2 r^2.$
In this case we have
\begin{eqnarray}
a(l) &=& {1 \over \sqrt{2} }\left( ip_r + {l+1 \over r} - \omega r\right), \\
\nu(l) &=& \left( l - {1\over 2} \right)\omega, \\
\epsilon(l) &=& \omega.
\end{eqnarray}
To get finite-norm eigenvectors one should assume the Case-II.
Putting $n=l_{max}$ we solve (\ref{caseii-a}) obtaining the following
normalized
eigenstate
\begin{equation}
\psi_{n,l_{max},m}(r\theta\phi) = \sqrt{ { \omega^{n+{3\over 2}} n! \over
\sqrt{\pi} (2n+1)!}
           }\,2^{n+1}\; r^{n} e^{-{\omega \over 2} r^2} \,
          Y_{l_{max},m}(\theta\phi).
\end{equation}
The eigenvalue of this state is $E_{n,l_{max}} = (n+{3\over 2})\,\omega.$
The normalized eigenstate for  $0\leq l <n$ is
\begin{equation}
\psi_{nlm} = \sqrt{ \omega^{l+{3\over 2}} n! \over \sqrt{\pi}(n-l)!(2n+1)!
}\,2^{l+1}\;
           e^{ {\omega\over 2} r^2} {1\over r^{l+1} }\left( {1\over r}{d\over
dr}\right)^{n-l}
           \left( r^{2n+1} e^{-\omega r^2} \right) \, Y_{lm},
\end{equation}
and the corresponging eigenvalue is
\begin{equation}
E_{nl} = (2n -l + {3\over 2})\,\omega.
\end{equation}
This way of solving the eigenvalue problem and the resulting forms of the wave
functions are
much nicer than the one from the series expansions and the Laguerre functions.

One may apply this idea to the eigenvalue problem of the hydrogen atom.  But it
is
rather trivial, and we turn to the spinning top problem whose corresponding
Hamiltonian is given
by (\ref{H_top_op}).  Since the eigenvalues of the Hermitian operators given by
(\ref{pp_op}) are  $p_\phi = m$ and $p_\chi = l$ with integer $m$ and $l,$
the Hilbert space ${\cal E}$ can be decomposed as
\begin{equation}
{\cal E} = \bigoplus_{l,m = -\infty}^\infty{\cal E}(l,m).
\end{equation}
The Hamiltonian in this subspace is
\begin{equation}
H(l,m)={p^\dagger_\theta p^{}_\theta \over 2I_1}
  + { (m-l\cos\theta)^2 \over 2I_1 \sin^2\theta}
  +{l^2 \over 2 I_3}.
\end{equation}
Then defining
\begin{eqnarray}
a(l,m) &=& {1\over \sqrt{2I_1}}\left(ip_\theta + {l\cos\theta -m \over
\sin\theta}\right), \\
\nu(l) &=& {1\over 2} \left( {l^2\over I_3} - {l\over I_1} \right),\\
\epsilon(l) &=& -{1\over 2} \left( {1\over I_3} - {1\over I_1}\right)(2l-1),
\end{eqnarray}
one can prove that the Hamiltonian operator $H(l,m)$ which acts in the subspace
${\cal E}(l,m)$ can be written as
\begin{eqnarray}
H(l,m) &=& a(l,m)^\dagger a(l,m) + \nu(l)\\
     &=& a(l+1,m)a(l+1,m)^\dagger + \nu(l+1) +\epsilon(l+1).
\end{eqnarray}
These show that the raising and lowering operations do not change the quantum
number $m.$
The normalized wave function $\psi_{n,l_{max},m}(\theta\phi\chi)$ for
$l=l_{max}\equiv n,$ which
can be solved from (\ref{caseii-a}), is
\begin{equation}
\psi_{n,l_{max},m}={1\over 4\pi}\sqrt{ {2\,(2n+1)! \over (n+m)!\, (n-m)!} }\;
         \left(\cos{\theta\over 2}\right)^{n+m}
         \left(\sin{\theta \over 2}\right)^{n-m} \, e^{in\chi + im\phi}.
\end{equation}
By checking the termination point of the descension and the behavior of the
wave function at
$\theta = 0$ and $\pi,$ it is not difficult to
show that
\begin{equation}
-l_{min} = l_{max} \geq |m|.
\end{equation}
The energy eigenvalue for general $l$ and $m$ is
\begin{equation}
E_{nl} = {n(n+1) \over 2I_1}
               + \left({1\over I_3}-{1\over I_1}\right) {l^2\over 2},
\end{equation}
which is in good agreement with known result \cite{fowler,landau}.
The wave function for this energy level can be obtained from (\ref{ket2}).
\vspace{.10in}

{\large\bf V. Conclusion}

Non-Hermitian representations of the canonical commutation
relations, rather than Hermitian ones, are much more useful for both
quantizing classical systems and solving eigenvalue equations.  To make it
clear
we introduced quantum canonical transformations of the second type with a
special interest
on those relations to the generalized ladder operators.
The usual operator ordering problems, which are inevitable when there
are no indisputable principles, do not occure when one follows the mixed matrix
element
technique of the canonical transformations. This means that for a given
classical Hamiltonian
one may write the Hamiltonian operator immediatly.
Another advantage of the non-Hermitian realization is that it is possible to
introduce the
generalized ladder operators naturally which may greatly simplify for solving
eigenvalue equations.
It is certain that one can not solve {\sl all} the eigenvalue equations in this
way.
But it is quite probable that whenever there is a dynamical symmetry associated
with
a Hamiltonian, the relevant generalized ladder opeartors can be found.

Applications to quantum field theories are open when one uses the functional
Schr\"odinger
equation formulations for these.  Our further investigation is aimed on this
project.
\vspace{.10in}

{\large\bf Acknowledgments}

This is supported by the Basic Science Research Institute Program of the
Ministry of
Education, Korea, BSRI-94-2436.

\newpage
{\large \bf APPENDIX: The eigenvalue problems of the angular momentum operator
using the generalized ladder operators}\vspace{.15in}
\renewcommand{\theequation}{A\arabic{equation}}
\setcounter{equation}{0}

Consider the following angular momentum operator
\begin{equation}
{\bf L}^2 = p^\dagger_\theta p^{}_\theta +\csc^2\theta\, p^2_\phi,
\end{equation}
where $0\leq \theta < \pi$ and $0\leq \phi <2\pi$ and
\begin{equation}
p_\theta^{\dagger} = {1\over \sin\theta}\, p^{}_\theta \sin\theta.
\end{equation}
The eigenvalues of the Hermitian $p_\phi$ are integers which are denoted
generally as $m.$
We decompose the Hilbert space ${\cal E}$ of ${\bf L}^2$ in terms of the
eigenvector spaces
${\cal E}(m)$ of $p_\phi$,
\begin{equation}
{\cal E} =\bigoplus_{m = -\infty}^{\infty}{\cal E}(m),
\end{equation}
in such a way that in each subspace ${\bf L}^2$ becomes
\begin{equation}
{\bf L}^2(m) =  p^\dagger_\theta p^{}_\theta +m^2\cot^2\theta +m^2.
\end{equation}
Here we used the fact $\csc^2\theta = \cot^2\theta + 1.$  It is easy to show
that when one
defines $a(m)$ as
\begin{equation}
a(m)=ip_\theta + m\cot\theta,
\end{equation}
it is in fact a generalized descending ladder opeartor.  The associated
relevant quantities
are
\begin{eqnarray}
\nu(m) &=& m(m-1),\\
\epsilon(m) &=& 0.
\end{eqnarray}
The normalized eigenstate for $l=m_{max},$ which can be solved from
(\ref{caseii-a}), is
\begin{equation}
\psi_{l,m_{max}}(\theta,\phi) = (-)^l{1\over 2^ll!}\sqrt{ (2l+1)! \over 4\pi}
                               e^{il\phi}\sin^l\theta, \;\; l=0,1,2,\dots
\end{equation}
By the action of $a(l)$ on this state the eigenvalue of ${\bf L}^2$ does not
change.
But for $p_\phi$ it changes from $l$ to $l-1.$ This process of descension
terminates at
$m_{min} = -l.$   That is, our string of descending states exactly coincides
with $Y_{lm},$
the spherical harmonics.

\end{document}